\documentclass[runningheads]{llncs}
\usepackage[T1]{fontenc}
\usepackage{graphicx}
\usepackage{hyperref}
\usepackage{cleveref}
\usepackage{microtype}
\usepackage[usenames,dvipsnames]{xcolor}
\usepackage{siunitx}
\usepackage{tikz}
\usepackage{color}

\usepackage{layouts}
\usepackage{multirow}
\usepackage{makecell}

\newcommand{\age}{\mathrm{Age}}
\newcommand{\sex}{\mathrm{Sex}}
\newcommand{\AD}{\mathrm{AD}}
\newcommand{\HCV}{\mathrm{HCV}}
\newcommand{\ECV}{\mathrm{ECV}}
\newcommand{\ICV}{\mathrm{ICV}}

\begin{document}
\title{Feature robustness and sex differences in medical imaging: a case study in MRI-based Alzheimer's disease detection}
\titlerunning{Feature robustness and sex differences in MRI-based Alzheimer's detection}
\author{Eike Petersen\inst{1}\orcidID{0000-0003-0097-3868} \and
Aasa Feragen\inst{1}\orcidID{0000-0002-9945-981X} \and
Maria Luise da Costa Zemsch\inst{1} \and
Anders Henriksen\inst{1} \and
Oskar Eiler Wiese Christensen\inst{1} \and
Melanie Ganz\inst{2,3}\orcidID{0000-0002-9120-8098}
for the Alzheimers Disease
Neuroimaging Initiative\thanks{Data used in preparation of this article was obtained from the Alzheimers Disease Neuroimaging Initiative (ADNI) database (\url{http://www.adni-info.org/}). The investigators within the ADNI contributed to the design and implementation of ADNI and/or provided data, but did not participate in analysis or writing of this report.}}
%
% index{Petersen, Eike}
% index{Feragen, Aasa}
% index{da Costa Zemsch, Maria Luise}
% index{Henriksen, Anders}
% index{Wiese Christensen, Oskar Eiler}
% index{Ganz, Melanie}

\authorrunning{E. Petersen et al.}
\institute{Technical University of Denmark, DTU Compute, Kgs. Lyngby, Denmark \and
University of Copenhagen, Department for Computer Science, Copenhagen, Denmark \and
Rigshospitalet, Neurobiology Research Unit, Copenhagen, Denmark \\
\email{ewipe@dtu.dk}, \email{afhar@dtu.dk}, \email{melanie.ganz@nru.dk}
}
\maketitle              % typeset the header of the contribution
\begin{abstract}
Convolutional neural networks have enabled significant improvements in medical image-based diagnosis.
It is, however, increasingly clear that these models are susceptible to performance degradation when facing spurious correlations and dataset shift, leading, e.g., to underperformance on underrepresented patient groups.
In this paper, we compare two classification schemes on the ADNI MRI dataset: a simple logistic regression model using manually selected volumetric features, and a convolutional neural network trained on 3D MRI data.
We assess the robustness of the trained models in the face of varying dataset splits, training set sex composition, and stage of disease. %, by applying a model trained on healthy controls and Alzheimer's disease patients to discriminate between these groups as well as between patients with stable and progressive mild cognitive impairment.
In contrast to earlier work in other imaging modalities, we do not observe a clear pattern of improved model performance for the majority group in the training dataset.
Instead, while logistic regression is fully robust to dataset composition, we find that CNN performance is generally improved for both male and female subjects when including more female subjects in the training dataset.
We hypothesize that this might be due to inherent differences in the pathology of the two sexes.
%In contrast to earlier work on diagnosing lung diseases based on chest x-ray data, we do not find a strong dependence of model performance for male and female test subjects on the sex composition of the training dataset.
Moreover, in our analysis, the logistic regression model outperforms the 3D CNN, emphasizing the utility of manual feature specification based on prior knowledge, and the need for more robust automatic feature selection.
\end{abstract}
\setcounter{footnote}{0}

\section{Introduction}
In recent years, various groups have reported highly accurate detection of Alzheimer's disease (AD) and progressive mild cognitive impairment (pMCI) -- which represents an earlier disease stage that continues to progress into AD or other types of dementia~\cite{Mielke2014} -- based on magnetic resonance imaging (MRI) volumes using convolutional neural networks (CNNs)~\cite{Wen2020}. %\cite{Moscoso2019,Tinauer2021,Basaia2019,Dai2021,Moscoso2019}.
Simultaneously, the potential brittleness of deep learning has become apparent due to issues like model underspecification~\cite{DAmour2020}, spurious correlations~\cite{Geirhos2020}, and susceptibility to dataset shift~\cite{Azulay2019}.
Multiple reviews in the medical domain have shown that deep learning-based publications often suffer from inadequate model reporting and overly optimistic performance estimates~\cite{Jacobucci2021,Wynants2020}. Wen et al.~\cite{Wen2020} found in their systematic review that half of the studies reporting on MRI-based AD detection using deep learning potentially suffered from data leakage, likely leading to inflated performance estimates.
Additionally, in clinical applications with low-dimensional input spaces, several systematic reviews have found no performance benefit of machine learning-based techniques over simple logistic regression~\cite{Cowling2021,Nusinovici2020}, raising the question of whether simple models based on manually extracted, low-dimensional features may also perform competitively in medical imaging. %Christodoulou2019

In a parallel development, the question of sex and gender-related performance disparities of machine learning models has lately received a lot of attention~\cite{Obermeyer2019,SeyyedKalantari2021}.
Studies have shown that CNNs can accurately identify a patient's age, sex, and ethnicity from chest x-ray images~\cite{Banerjee2021,Yi2021}, and that chest x-ray classifiers tend to underdiagnose underserved patient populations~\cite{SeyyedKalantari2021}.
Larrazabal et al.~\cite{Larrazabal2020} have analyzed the effect of training dataset sex imbalance on a chest x-ray classifier's performance for male and female subgroups, finding a consistent decrease in performance for the underrepresented sex.
To the authors' knowledge, similar analyses have yet to be performed in the brain MRI classification setting, even though it is known that AD presents differently in males and females~\cite{Mielke2014,Podcasy2016}.

\begin{figure}[b]
    \includegraphics[width=\textwidth]{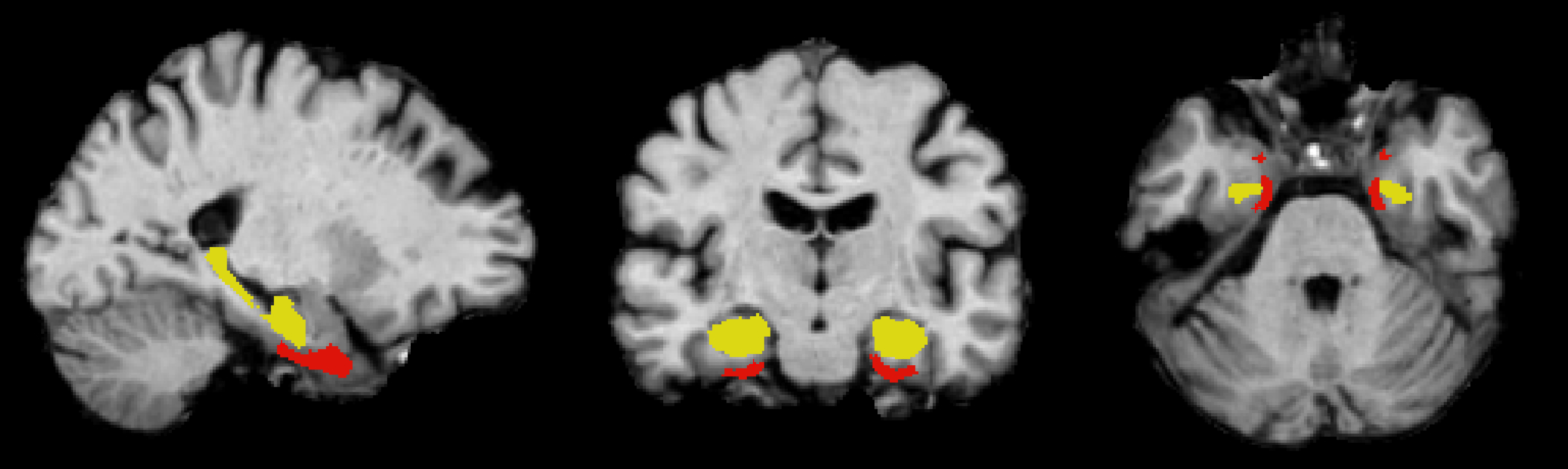}
    \caption{Slices of an exemplary recording used in this study, skull-stripped and registered to a common space. Hippocampus and entorhinal cortex (yellow and red, segmented using FreeSurfer) are highlighted. Left to right: sagittal, coronal, and axial view.}
    \label{fig:MRI-ex}
\end{figure}

Training on a sex-imbalanced dataset and then evaluating model performance on the underrepresented group represents a particular type of \emph{dataset shift}~\cite{QuinoneroCandela2009}, and the close connection between model robustness (to dataset shift and other challenges) and algorithmic fairness has been emphasized recently~\cite{Adragna2020}.
To achieve robustness to dataset shifts, the \emph{feature space} in which classification is performed plays a crucial role: to be robust against dataset shifts, estimation must be performed in a feature space that gives rise to a classifier that is optimal across different environments~\cite{Arjovsky2019}.
This feature space can either be manually crafted or automatically inferred, as is usually the case in deep learning~\cite{Abrol2021}.

In this paper, we analyze the robustness of two different MRI-based classifiers to distribution shifts.
Both classifiers are trained to detect Alzheimer's disease, based on different feature representations.
The first is a simple logistic regression model using manually selected volumetric features, which are obtained using standard MRI processing tools~\cite{ashburner2012spm,fischl2012freesurfer}.
The second is a CNN using full 3D MRI volumes.
Both models are considered state-of-the-art for Alzheimer's disease classification.
For analyzing the classifiers' robustness, we consider two separate types of distribution shifts:
Firstly, we analyze the effect of differing training dataset sex and age compositions on the performance for male/female or young/old test subjects, similar to the analysis of Larrazabal et al.~\cite{Larrazabal2020}.
And secondly, like various other groups have done~\cite{Wen2020} (although not combined with the sex and age imbalance analyses we perform), we evaluate the performance of classifiers trained on subjects diagnosed with AD and healthy controls on a test set consisting of subjects with stable and progressive MCI.
%Finally, we evaluate the influence of random dataset splits.

\section{Methods}
\subsection{Dataset}
\begin{table}[t]
    \renewcommand{\arraystretch}{1.2}
    \setlength{\tabcolsep}{5pt}
    \caption{Composition of the dataset used in this study (based on the ADNI dataset~\cite{Jack2008}), stratified by sex and field strength.}
    \centering
    \begin{tabular}{ c c c c c c } 
     Diagnosis & Male & Female & $\SI{1.5}{\tesla}$ & $\SI{3}{\tesla}$ & Total \\ 
     \hline
     AD & 181 (54.03\%) & 154 (45.97\%) & 175 (52.24\%) & 160 (47.76\%) & 335 \\
     HC & 282 (43.52\%) & 366 (56.48\%) & 219 (33.80\%) & 547 (66.20\%) & 648 \\
     pMCI & 149 (58.66\%) & 105 (41.34\%) & 184 (72.44\%) & 70 (27.56\%) & 254 \\
     sMCI & 155 (59.62\%) & 105 (40.38\%) & 122 (46.92\%) & 138 (53.08\%) & 260 \\
    \end{tabular}
    \label{tab:data}
\end{table} % to get these numbers: open python, instantiate ADNI_Feature_Module and call the two load_df functions
We use MRI volumes from the ADNI dataset~\cite{Jack2008} (700, 463, and 334 recordings from ADNI1, ADNI2, and ADNI3, respectively), including a single T1-weighted structural MRI volume (acquired on different MR scanners using a field strength of either $\SI{1.5}{\tesla}$ or $\SI{3}{\tesla}$) per subject in our experiments.
Like various other studies~\cite{Wen2020}, we consider two groups of subjects, giving rise to two classification tasks:
firstly, healthy control (HC) and Alzheimer's disease (AD) subjects and, secondly, stable and progressive mild cognitive impairment (sMCI, pMCI) subjects.
Subjects were labeled as pMCI in our analysis if they were diagnosed with MCI at the time of the recording and then were diagnosed with AD at any point during the following five years.
(This definition only encompasses (progressive) \emph{amnestic} MCI, since non-amnestic MCI may develop into non-AD dementias~\cite{Mielke2014}.)
Subjects were labeled as sMCI if they were diagnosed with MCI at the time of the recording, not diagnosed with AD at any point during the five-year follow-up period, and if there was at least one valid diagnosis from years three to five after the initial recording was made.
The dataset is summarized in \cref{tab:data}.
The median age across the whole dataset was $73.0$ years; the average age of the different subject cohorts was $74.99\pm 7.08$ (males), $72.70\pm 6.94$ (females), $75.13\pm 7.83$ (AD/pMCI cases), $73.08 \pm 6.58$ (HC/sMCI cases), $75.70 \pm 7.91$ (male AD/pMCI cases), and $71.96 \pm 6.46$ (female AD/pMCI cases).

\subsection{AD Classification using logistic regression}
%\subsubsection{Preprocessing.}
Intracranial volume (ICV) was quantified as the combined volumes of gray- and white matter and cerebrospinal fluid~\cite{Malone2015}, segmented with SPM12~\cite{ashburner2012spm}.
Hippocampal volume (HCV) and entorhinal cortex volume (ECV) was extracted from structural MRI volumes  using FreeSurfer v7.1.1~\cite{fischl2012freesurfer}, see~\cref{fig:MRI-ex} for an example.

%\subsubsection{Classifier}
A simple logistic regression (LR) model of the form
\begin{equation}
    P(\AD) \approx q_{\AD} = \sigma(\theta_1 \cdot \age + \theta_2 \cdot \ICV + \theta_3 \cdot \HCV + \theta_4 \cdot \ECV),
\end{equation}
with $\sigma(\cdot)$ denoting the sigmoid function, was fitted using stochastic gradient descent for 4000 epochs (initial learning rate $10^{-3}$, automatic learning rate scheduling -- if there is no improvement in the validation loss for 10 epochs, the learning rate is halved --, momentum~$0.9$, early stopping based on the validation loss, batch size 256).
The loss function was binary cross-entropy with $L_2$ regularization (regularization constant $10^{-4}$).

\subsection{AD Classification using CNNs}
%\subsubsection{Preprocessing}
The subject-specific MRI images were skull-stripped and registered to a common space (MNI305) using FreeSurfer v7.1.1~\cite{fischl2012freesurfer}, resulting in spatially normalized grayscale volumes of the dimensionality $256 \times 256 \times 256\, \si{mm^3}$.
These were then cropped to include the whole brain ($186 \times 186 \times 191\, \si{mm^3}$).
%\subsubsection{Dataset augmentation}
Dataset augmentation was performed using torchio~\cite{PerezGarcia2021}: During training, 80\% of the training samples were transformed using one (randomly selected) of the following transformations: rotation, elastic deformation, flipping, blurring, addition of Gaussian noise, and addition of an MRI bias field, spike, ghosting, or motion artifact.

%\subsubsection{Classifier} 
A convolutional neural network (CNN) was trained, using the preprocessed 3D MRI volumes as inputs and the subject's AD/HC label as the target.
The model architecture was inspired by Tinauer et al.~\cite{Tinauer2021} and Wen et al.~\cite{Wen2020}, and was (manually) selected based on validation set performance.
The convolutional part of the model consists of eight consecutive convolution layers (all with sixteen channels), each followed by a rectified linear unit (ReLu).
The first two layers use kernels of size $5\times 5\times 5$; the subsequent six layers use $3 \times 3 \times 3$ kernels.
Every second layer uses a stride of two, thus progressively reducing image resolution.
This is followed by a dropout layer ($p=0.5$), a fully connected linear layer with 32 output channels, a ReLu, another dropout layer ($p=0.5$), and the final, fully connected classification layer with sigmoid activation.
The model has 337,000 trainable parameters, which are estimated using the Adam optimization algorithm (initial learning rate $10^{-4}$, automatic learning rate scheduling based on the validation loss as described above, 200 epochs, batch size 6, early stopping based on the validation loss) to minimize binary cross-entropy (no regularization).

\subsection{Performance and robustness evaluation}
\begin{figure}[t]
    \centering
    \includegraphics[width=\textwidth]{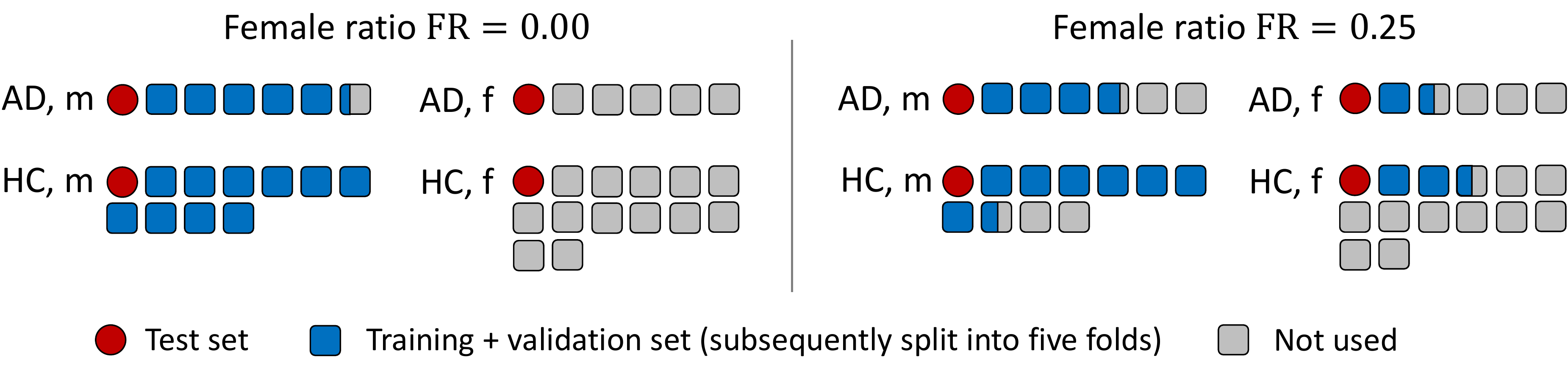}
    \caption{Visualization of the AD/HC dataset and two exemplary splits. Each marker represents 25 subjects (rounded). For clarity, only one of the five test sets and only two of the five training and validation datasets for this test set (all with different male--female ratios $\mathrm{FR}=\frac{n_{\!f}}{n_{\!f}+n_m}$, with $n_{\!f}$ and $n_m$ the number of females and males in the combined training and validation dataset) are shown.
    %Training and validation datasets are always of size 379 with a ratio of $\SI{34.1}{\percent}$ AD cases.
 %   Not shown here: each combined training and validation dataset is split into five folds, one of which is selected for validation, while the other four make up the training dataset.
    }
    \label{fig:data}
\end{figure}
To evaluate the two models' overall performance and robustness, we trained them on 125 (LR) and 50 (CNN) different training datasets and evaluated them on six different test sets.
First, five AD/HC test sets of size 100 were drawn from the full dataset with 25 samples each of male/female AD/HC subjects, with no subject overlap between the test sets.
For each of the five test sets, five training and validation sets of combined size 379 were drawn from the remaining AD/HC subjects (without replacement), with five different male-to-female ratios ($\SI{0}{\percent}$ to $\SI{100}{\percent}$ females in $\SI{25}{\percent}$ increments) and a constant fraction of $\SI{34.1}{\percent}$ AD cases, corresponding to the fraction of AD cases in the whole dataset.
The gender-specific datasets were nested: female subjects used at a female ratio (FR) of $\SI{50}{\percent}$ represent a subset of female subjects used at an FR of $\SI{75}{\percent}$, etc.

Each of the 25 combined training and validation sets was then split randomly into five folds following standard recommendations~\cite{varoquaux2017assessing}.
For the logistic regression, each fold was used once as the validation set, resulting in a total of five training--validation set combinations per test set (of which there are five) and sex ratio (of which there are five), yielding a total of 125 different training and validation datasets.
For the CNNs, due to the computational effort, only the first two of the five fold combinations were used per sex ratio, thus yielding a total of 50 different training and validation datasets. %(five training datasets with different sex ratios for each of the five different test sets).
Each model was then evaluated on both the respective AD/HC test set and the full sMCI/pMCI dataset (which had not been used for training), calculating performance metrics on both males and females.
\Cref{fig:data} illustrates the full AD/HC dataset and two exemplary data splits.

The decision thresholds for all models were selected to maximize the geometric mean of sensitivity and specificity on the validation dataset~\cite{Fernandez2018b}. %Kubat1997
All models were implemented and trained using PyTorch Lightning v.1.5.9~\cite{Falcon2019} and training one CNN took about twelve hours on a single NVIDIA Titan~X GPU.
When the performance of models trained on the same datasets was compared, statistical significance was assessed using Wilcoxon signed-rank tests.
 %and two Intel Xeon E5-2620 CPUs (partially shared with other users).
%Image augmentation, which was performed on the CPU, represented a performance bottleneck.

An analogous analysis was also performed with respect to age groups instead of sex; its details and results can be found in the supplementary material.

\section{Results}
\begin{figure}[p]
    \centering
    \begin{tikzpicture}
        \node[below right, inner sep=0](image) at (0,0) {
            \includegraphics{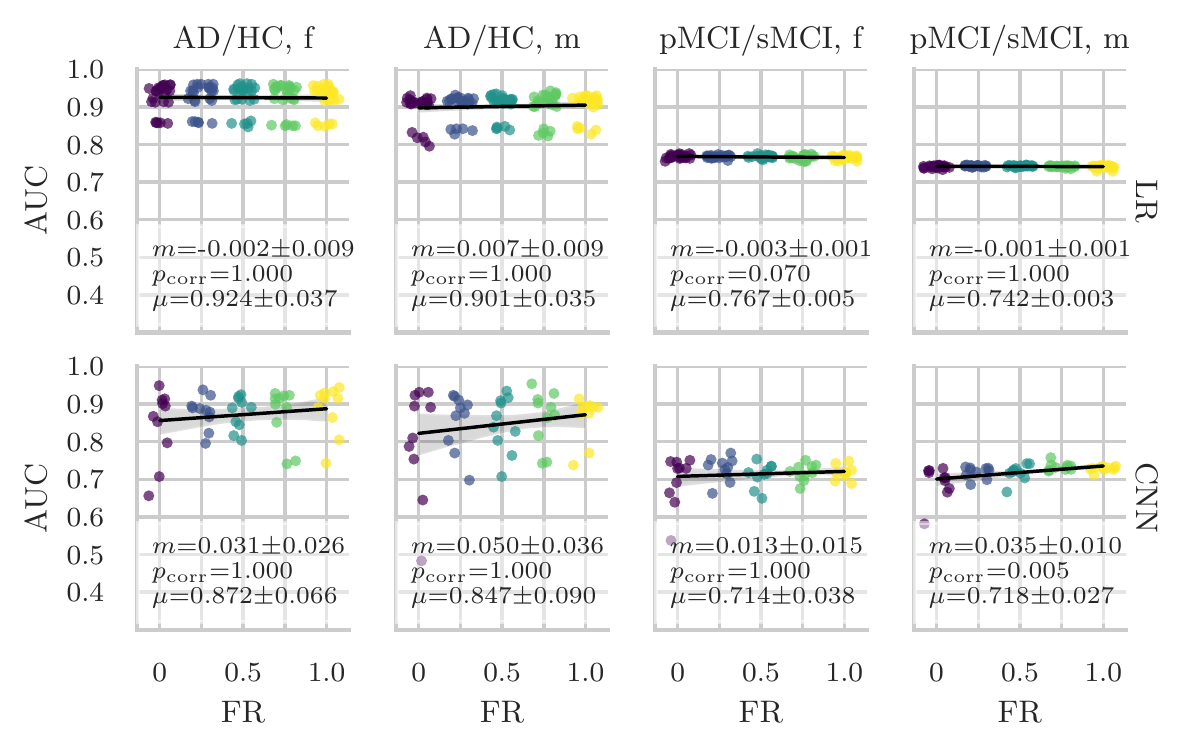}
        };
        \node[] (0, 0) {\textbf{(A)}};
    \end{tikzpicture}    
    
    \vspace{-4pt}
    
    \begin{tikzpicture}
        \node[below right, inner sep=0](image) at (0,0) {
            \includegraphics{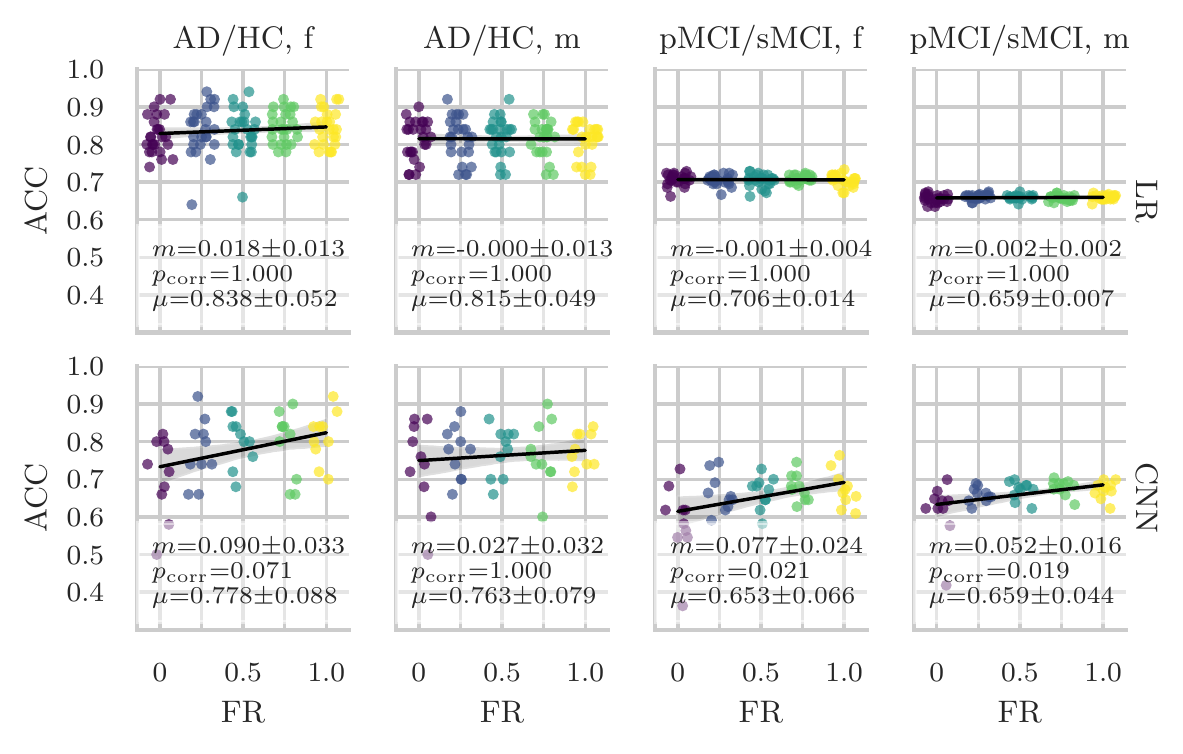}
        };
        \node[] (0, 0) {\textbf{(B)}};
    \end{tikzpicture}  
    
    \caption{Distribution of (A) the area under the curve (AUC) of the receiver-operating characteristic and (B) the accuracy achieved by the trained models (first row in both panels: LR, second row: CNN) on the different test sets. FR: ratio of female subjects in the training and validation set, $m$: slope of the regression line ($\pm$ standard deviation), Bonferroni corrected p-value null hypothesis: $m=0$, $\mu$: average AUC / ACC ($\pm$ standard deviation) across all sex ratios. Random jitter is added to all x coordinates to prevent excessive overlap -- all points are sampled at $\mathrm{FR} \in \{0, 0.25, 0.5, 0.75, 1.0\}$, as indicated by the five different colors. Note the truncation of all y-axes.}
    \label{fig:aucs}
\end{figure}
%\begin{figure}[t]
%    \centering
%    \includegraphics{analysis3-acc.pdf}
%    \caption{Distribution of the accuracy (ACC) achieved by the trained models on the different test sets are shown in the top row for LR and bottom row for the CNN. See the caption of \cref{fig:aucs} for further details.}
%    \label{fig:accs}
%\end{figure}
\Cref{fig:aucs} shows the distribution of the area under the curve (AUC) of the receiver-operating characteristic and the accuracy (ACC) achieved by the trained models on the different test sets.
To assess the dependence of model performance for males and females on the training dataset sex composition, regression lines were fit and a t-test was performed to assess whether the slope was significantly different from zero.
To account for multiple comparisons, a Bonferroni correction with a factor of eight was performed.
The figure reports the corrected p-values.

%and \cref{tab:results} summarizes the results.
%\begin{table}[t]
%    \renewcommand{\arraystretch}{1.5}
%    \setlength{\tabcolsep}{5pt}
%    \caption{Mean and standard deviation of trained model performance across all training and testing conditions. AUC: Area under the curve of the receiver-operating characteristic, ACC: classification accuracy. Significance of male--female differences assessed using a paired t-test.}
%    \centering
%    \begin{tabular}{c c c c c c} 
%     Model & Metric & Test set & male & female & p-value\\ 
%     \hline
%        \multirow{4}{*}{LR} & \multirow{2}{*}{AUC} & AD/HC & $0.899 \pm 0.033$ & $0.923 \pm 0.038$ & \\
%         & & pMCI/sMCI & $0.743 \pm 0.003$ & $0.766 \pm 0.005$ & \\      
%         & \multirow{2}{*}{ACC} & AD/HC & $0.815 \pm 0.043$ & $0.841 \pm 0.049$ & \\ 
%         & & pMCI/sMCI & $0.661 \pm 0.009$ & $0.704 \pm 0.014$ & \\          
%      \hline
%        \multirow{4}{*}{CNN} & \multirow{2}{*}{AUC} & AD/HC & $0.90 \pm 0.03$ & $0.92 \pm 0.04$ & \\
%         & & pMCI/sMCI & & & \\      
%         & \multirow{2}{*}{ACC} & AD/HC & $0.81 \pm 0.04$ & $0.84 \pm 0.05$ & \\ 
%         & & pMCI/sMCI & & & \\      
%    \end{tabular}
%    \label{tab:results}
%\end{table}
As also found in previous studies~\cite{Wen2020}, both models drop in  performance on sMCI~vs.~pMCI classification compared to HC~vs.~AD classification, for which they were trained.
%\footnote{This performance gap may be partially explained by our (and other studies') definition of pMCI, which only encompasses progression to AD and neglects progression to non-AD dementias.}
The LR model achieves consistently high AUC and ACC across all training and validation sets, and performance is similar for male and female test subjects, although significantly better for females ($p_{\textrm{corr}}=\num{5.5e-4}$ and $p_{\textrm{corr}}=\num{5.9e-22}$ for AUC on AD/HC and pMCI/sMCI, respectively, with Bonferroni correction factor~$2$).
For the LR model, all regression slopes were non-significant ($p_{\textrm{corr}}\gg0.05$), indicating no significant effect of dataset composition on model performance.
The CNN performed significantly worse compared to the LR ($p_{\textrm{corr}}=\num{2.8e-25}$ for AUC, $p_{\textrm{corr}}=\num{6.2e-12}$ for ACC, with Bonferroni correction factor~$2$), and it exhibited a stronger dependence of performance for male and female test subjects on the training dataset composition.
Three statistically significant positive correlations were observed: between FR and AUC in male pMCI/sMCI subjects ($p_{\textrm{corr}}=\num{5.4e-3}$), and between FR and ACC in female ($p_{\textrm{corr}}=\num{2.1e-2}$) and male ($p_{\textrm{corr}}=\num{1.9e-2}$) pMCI/sMCI subjects.

%(as indicated by larger slope values~$m$ in \cref{fig:aucs}), although in all cases except male pMCI/sMCI subjects this was not statistically significant ($p \gg 0.05$).
%The observed (although weak and not statistically significant) trends agree with previously reported findings reporting improved performance on female test subjects for training on mostly female data, and vice versa.

\section{Discussion \& conclusion}
In the present study, motivated by recent work demonstrating performance disparities in medical image classifiers between patient populations~\cite{Larrazabal2020,SeyyedKalantari2021}, we have analyzed the robustness of two state-of-the-art brain MRI classifiers and their associated feature representations to multiple types of distribution shift. Logistic regression using manually extracted volumetric features was compared to a standard 3D CNN classifier using the full 3D MRI volumes as inputs.
Both models performed within the range reported for state-of-the-art CNN-based AD classification in the review of Wen et al.~\cite{Wen2020} for studies without suspected data leakage.
The (valid) studies that report higher performance either use significantly more recordings or multiple modalities, and they only report the performance of a single training run.
(Some of our runs performed significantly better than the mean accuracy reported above, see \cref{fig:aucs}.)
The LR model, using manually selected volumetric features, performed significantly better on average than the 3D CNN using the full MRI images.
While this might change for larger dataset sizes, our sample size is comparable to typical sample sizes used in this domain.

We analyzed the effect of different ratios of male and female examples in the training dataset on the performance of the trained classifier for male and female test subjects. %The robustness of both classifiers to two types of distribution shift (shifts in the population sex ratio and in the disease stage) and random dataset splits was assessed.
In our study, only the CNN's performance showed a dependence on training dataset composition, while the LR model was robust to this variation.
This is not surprising, given that the LR model is based on a given representation (the extracted brain volumes and subject age), whereas the CNN performs representation learning and is, thus, prone to tailor the learned representation specifically to the training dataset.
An important avenue for future research concerns the question of whether environment-invariant feature representations can be inferred automatically; an end towards which various methods have been proposed~\cite{Arjovsky2019,Pawlowski2020,Zhao2020}.
This would potentially allow for mitigating the effect of training dataset composition on male and female test subject performance.
Generalization to different populations remains an unsolved problem in brain MRI AD analysis~\cite{Wen2020}, with important consequences for the general robustness and fairness of the resulting systems. Finally, as a general trend, we observe that the influence of training dataset composition on accuracy seems to be larger than the influence on AUC, as indicated by larger slopes in \cref{fig:aucs}.
This appears plausible, given that the accuracy is influenced by one additional, dataset-dependent processing step compared to AUC: threshold selection.

Interestingly, we find that CNN performance (as measured by AUC and ACC) is improved in both men and women by including more women in the training dataset.
This differs from the results of our age splitting experiment, where, while most relationships are not statistically significant, we observe performance on younger subjects to improve the more younger subjects are in the training dataset, whereas performance on older subjects tends to decline.
These contrasting results underline the complexity of the relationship between dataset composition and subgroup performance.
We hypothesize that the beneficial influence of female subjects in the training dataset on model performance on male test subjects may be explained by sex differences in pathological severity between AD study subjects.
Several epidemiological studies have shown that neurodegeneration and clinical symptoms occur more rapidly for females once a diagnosis is suspected~\cite{Podcasy2016}.
Moreover, women might present stronger cognitive decline than men with the same level of brain pathology~\cite{Mielke2014}, thus being diagnosed with AD at an earlier stage of disease progression than men.
This might render female subjects more helpful as training examples to a CNN classifier compared to male subjects, and it would also explain why we generally observe better performance for females subjects.
More research is certainly required to further investigate and explain these observations, however.

%we observed a weaker dependence of classifier performance for male and female test subjects on the training dataset sex composition than observed in a previous study on lung disease detection from chest x-ray data~\cite{Larrazabal2020}.
%The small effect of the training dataset sex composition observed in our study is in line with previous research showing sex-related differences in MRI recordings to be limited and gradual~\cite{zhang2021human}, as opposed to the large inter-sex differences in chest x-ray recordings.

%Nevertheless, it is an important result that -- despite known sex differences in AD~\cite{Mielke2014} -- the training dataset composition does not appear to have a strong effect on male and female test subject performance, indicating that the employed feature representations are -- to some degree, at least -- invariant to this dataset shift~\cite{Arjovsky2019}.

One potential limitation of our study concerns the preprocessing employed for both models.
The employed segmentation steps are partially based on MRI atlases, which have been extracted from study databases.
Thus, even in the cases in which the training dataset nominally contained no males or no females, this preprocessing step still incorporated some information about male and female subjects.
The degree to which this influenced our analyses is challenging to quantify, but it represents a potentially confounding factor that might cause us to underestimate the effect of the training dataset composition.
However, practical applications also typically employ atlas-based preprocessing, making our analysis closer to practical scenarios than if we had omitted these preprocessing steps.

%{\color{BrickRed} This whole section is unfinished (because the results are not final and may still change significantly.} Feature selection has a strong impact on the robustness of the whole machine learning process and generalization capability... either manually craft useful features (if prior knowledge available and resulting performance sufficient) or make really sure that DNN learns useful features by using very large and representative datasets, clever feature learning stuff, and/or actually analyzing the learned features using XAI methods?

%Discuss how well our models perform compared to previously reported results~\cite{Wen2020,Basaia2019,Moscoso2019,Tinauer2021}. 
%Wen et al. find ``On the ADNI test dataset, the diagnostic BA of CNNs ranged from 0.76 to 0.89 for the AD vs CN task and from 0.69 to 0.74 for the sMCI vs pMCI task... In the current study, the SVM was at least as good as the best CNNs for both the AD vs CN and the sMCI vs pMCI task... Overall, these results bring important
%information. First, good generalization to unseen, similar, subjects
%demonstrate that the models did not overfit the subjects at hand
%in the training/validation set. On the other hand, poor generaliza-
%tion to different age ranges, protocols and inclusion criteria show
%that trained models are too specific of these characteristics. Gen-
%eralization across different populations thus remains an unsolved
%problem and will require training on more representative datasets
%but maybe also new strategies to make training more robust to
%heterogeneity.''

\subsubsection*{Code \& data availability}
The full code for all implemented models, dataset splitting, and statistical analyses is available online in our GitHub repository: \url{https://github.com/e-pet/adni-bias}.

\subsubsection*{Acknowledgements}
We thank Morten Rieger Hannemose for helpful comments on the manuscript and the statistical analysis.
This research was supported by Danmarks Frie Forskningsfond (9131-00097B), the Novo Nordisk Foundation through the Center for Basic Machine Learning Research in Life Science (NNF20OC0062606) and the Pioneer Centre for AI, DNRF grant number P1.
%The authors would like to thank Morten Hannemose, DTU Compute, for helpful comments on the manuscript and the statistical analysis.
%This research was supported by Danmarks Frie Forskningsfond {(\color{BrickRed} GRANT NUMBER TBD), AND OTHER FUNDING?}.
Data collection and sharing for this project was funded by the ADNI (National Institutes of Health Grant U01 AG024904).
ADNI is funded by the National Institute on Aging, the National Institute of Biomedical Imaging and Bioengineering, and through generous contributions from private sector institutions.
The Canadian Institutes of Health Research is providing funds to support ADNI clinical sites in Canada.
Private sector contributions are facilitated by the Foundation for the National Institutes of Health (www.fnih.org).
The grantee organization is the Northern California Institute for Research and Education, and the study is coordinated by the Alzheimer’s Disease Cooperative Study at the University of California, San Diego.
ADNI data are disseminated by the Laboratory for Neuro Imaging at the University of California, Los Angeles.

\bibliographystyle{splncs04}
\bibliography{references}

%Author, F.: Article title. Journal \textbf{2}(5), 99--110 (2016)

%Author, F., Author, S.: Title of a proceedings paper. In: Editor,
%F., Editor, S. (eds.) CONFERENCE 2016, LNCS, vol. 9999, pp. 1--13.
%Springer, Heidelberg (2016). \doi{10.10007/1234567890}

%Author, F., Author, S., Author, T.: Book title. 2nd edn. Publisher,

%Author, A.-B.: Contribution title. In: 9th International Proceedings
%on Proceedings, pp. 1--2. Publisher, Location (2010)

%LNCS Homepage, \url{http://www.springer.com/lncs}. Last accessed 4
%\end{thebibliography}

\newpage

\section*{Supplementary material}
\begin{figure}[p]
	\centering
	\begin{tikzpicture}
		\node[below right, inner sep=0](image) at (0,0) {
			\includegraphics{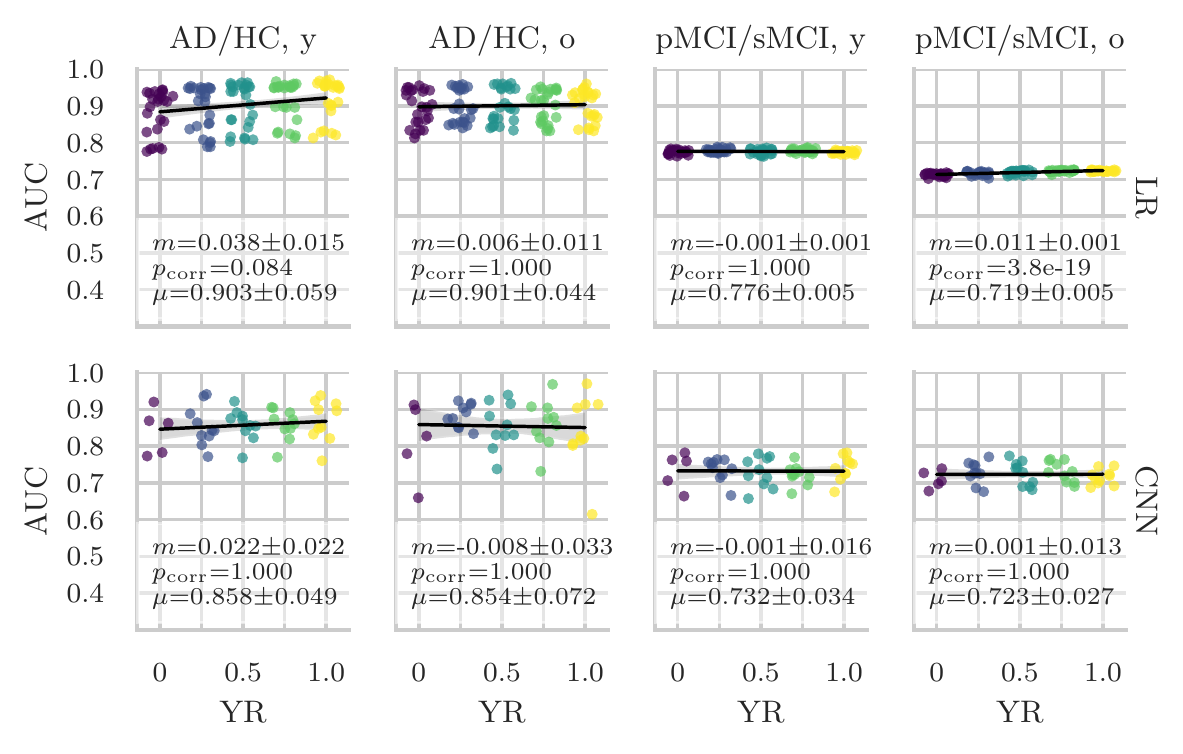}
		};
		\node[] (0, 0) {\textbf{(A)}};
	\end{tikzpicture}    
	
	\vspace{-4pt}
	
	\begin{tikzpicture}
		\node[below right, inner sep=0](image) at (0,0) {
			\includegraphics{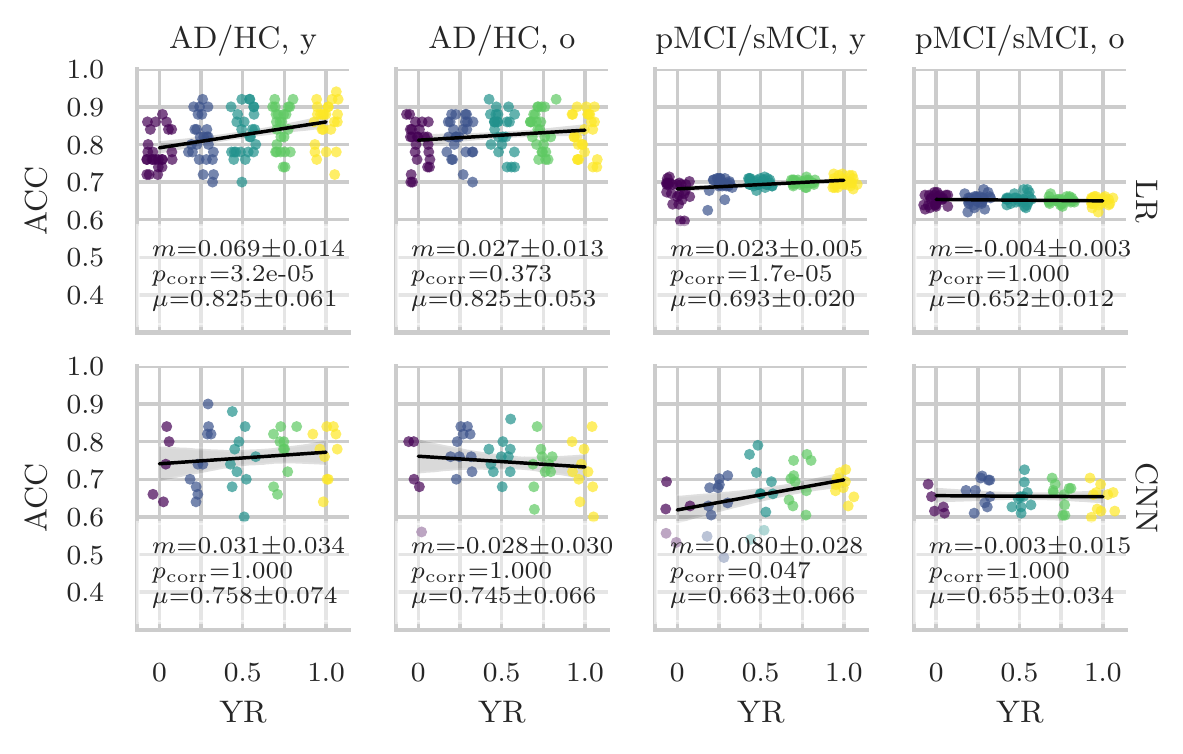}
		};
		\node[] (0, 0) {\textbf{(B)}};
	\end{tikzpicture}  
	
	\caption{Distribution of (A) the area under the curve (AUC) of the receiver-operating characteristic and (B) the accuracy achieved by the trained models (first row in both panels: LR, second row: CNN) on the different test sets. YR: ratio of younger ($\text{age} \leq 73.0$) subjects in the training and validation set, $m$: slope of the regression line ($\pm$ standard deviation), p-value null hypothesis: $m=0$, $\mu$: average AUC / ACC ($\pm$ standard deviation) across all younger/older ratios. Random jitter is added to all x coordinates to prevent excessive overlap -- all points are sampled at $\mathrm{YR} \in \{0, 0.25, 0.5, 0.75, 1.0\}$, as indicated by the five different colors. Note the truncation of all y-axes.}
	\label{fig:age-analysis}
\end{figure}

To analyze the influence of age, the same analysis as described in the main text was repeated, but splitting subjects into two age groups on the median age (group~y: $\text{age} \leq 73.0$, and group~o:~$\text{age} > 73.0$) instead of splitting based on sex.
In this case, the logistic regression model included sex as a variable, i.e., the model reads
\begin{equation}
	P(\AD) \approx q_{\AD} = \sigma(\theta_1 \cdot \age + \theta_2 \cdot \ICV + \theta_3 \cdot \HCV + \theta_4 \cdot \ECV + \theta_5 \cdot \sex).
\end{equation}
The results of this analysis can be found in \cref{fig:age-analysis}, where YR denotes the ratio of younger patients in the training dataset.
Only two relationships are both statistically significant and show a relevant effect size: the effect of YR on ACC in younger AD/HC subjects using LR ($p_{\textrm{corr}}=\num{3.2e-5}$, $m=0.069\pm 0.014$) and the effect of YR on ACC in younger pMCI/sMCI subjects using the CNN ($p_{\textrm{corr}}=\num{0.047}$, $m=0.080\pm 0.028$).
These relationships are as expected: the performance on young subjects improves as more young subjects are part of the training dataset.
%Significant positive correlations are observed between  (YR) and AUC on older pMCI/sMCI subjects using LR ($p_{\textrm{corr}}=\num{3.8e-19}$), ACC on younger AD/HC subjects using LR ($p_{\textrm{corr}}=\num{3.2e-5}$), ACC on younger pMCI/sMCI subjects using LR ($p_{\textrm{corr}}=\num{1.7e-5}$), and ACC on younger pMCI/sMCI subjects using a CNN ($p_{\textrm{corr}}=\num{4.7e-2}$).

\end{document}